\documentclass{ws-ijmpe}
\usepackage[super,compress]{cite}
\begin{document}

\markboth{Y. EL BASSEM $\&$ M. OULNE}{GROUND STATE PROPERTIES IN HFB METHOD}

\catchline{}{}{}{}{}

\title{GROUND STATE PROPERTIES OF EVEN-EVEN AND ODD Nd, Ce AND Sm ISOTOPES IN HARTREE-FOCK-BOGOLIUBOV METHOD}

\author{Y. EL BASSEM$^\dagger$ and M. OULNE\footnote{corresponding author.}}

\address{High Energy Physics and Astrophysics Laboratory, Department of Physics, \\Faculty of Sciences SEMLALIA, Cadi Ayyad University,  \\P.O.B. 2390, Marrakesh, Morocco.\\
$^\dagger$younes.elbassem@edu.uca.ma\\
$^*$oulne@uca.ma}

\maketitle

\begin{history}
\received{Day Month Year}
\revised{Day Month Year}
\end{history}

\begin{abstract}
In this work, we have studied ground-state properties of both even-even and odd $Nd$ isotopes within Hartree-Fock-Bogoliubov method with SLy5 Skyrme force in which the pairing strength has been generalized with a new proposed formula. We calculated binding energies, two-neutron separation energies, quadrupole deformation, charge, neutron and proton radii. Similar calculations have been carried out for Ce and Sm in order to verify the validity of our pairing strength formula. The results have been compared with available experimental data, the results of Hartree-Fock-Bogoliubov calculations based on the D1S Gogny effective nucleon-nucleon interaction and predictions of some nuclear models such as Finite Range Droplet Model (FRDM) and Relativistic Mean Field (RMF) theory.
\end{abstract}

\keywords{Hartree-Fock-Bogoliubov method; $Nd$, $Sm$ and $Ce$ isotopes; binding energy; proton, neutron and charge radii; two-neutron separation energies, quadrupole deformation.}

\ccode{PACS numbers: 21.10-k, 21.10.Dr, 21.10.Ft, 21.60-n}


\section{Introduction}
In nuclear structure theory, several approaches have been developed to study ground-state and single-particle (s.p) excited states properties of even-even and odd nuclei. Among them we can find ab-initio calculations (Green's function - Monte Carlo shell model) based on bare N-N interaction for the lightest nuclei \cite{Navratil}. For medium-mass nuclei up to A $\sim$ 60, the large-scale shell model \cite{Koonin} may be used. While for heavier nuclei, non-relativistic \cite{Terasaki,Dobaczewski,Chabanat,Stoitsov2000,Teran} and relativistic \cite{Ring96,Lalazissis} mean field theories are mostly used. The most popular one among them is the Hartree-Fock method or the Hartree-Fock + BCS in which the pairing correlations are added to the mean field via a corresponding potential term. Such methods give a good description of the nuclear structure near the line of $\beta$-stability \cite{Werner}. But, as one goes far from this line, the pairing correlations start to increase radically, so the HF+BCS theory ceases 
to be 
adequate for studying nuclei lying close to neutron and proton drip-lines. Thus, it is required to consider both the mean field and the pairing one self-consistently within the Hartree-Fock-Bogoliubov (HFB) Theory\cite{Yamagami}.

The aim of this work is to calculate and analyze some ground-state properties of even-even and odd $Nd$ isotopes using Skyrme-Hartree-Fock-Bogoliubov method and a new generalized formula for the pairing strength for a wide range of neutron numbers. The ground-state properties we have focused on are binding energy, two-neutron separation energy, charge, proton and neutron radii. We have also performed similar calculations for $Ce$ and $Sm$ which are in the vicinity of $Nd$.

The paper is organized as follows: in Section II, we briefly describe the Hartree-Fock-bogoliubov method . In Section III, some details about the numerical calculations are presented while in Section IV, we present our results and discussion. A conclusion is given in Section V.

\section{Hartree-Fock-Bogoliubov Method}
In Hartree-Fock-Bogoliubov method, a two-body Hamiltonian of a system of fermions can be expressed in terms of a set of annihilation and creation
operators $(c, c^\dagger)$:
\begin{equation}
 H=\sum_{n_1 n_2} e_{n_1 n_2} c_{n_1}^\dagger c_{n_2} + \frac{1}{4} \sum_{n_1 n_2 n_3 n_4} \bar{\nu}_{n_1 n_2 n_3 n_4} c_{n_1}^\dagger c_{n_2}^\dagger c_{n_4} c_{n_3}
 \label{eq1}
\end{equation}
with the first term corresponding to the kinetic energy and $\bar{\nu}_{n_1 n_2 n_3 n_4}=\langle n_1 n_2 | V | n_3 n_4 - n_4 n_3 \rangle$ are anti-symmetrized two-body interaction matrix-elements. \vspace{0.4em}
So, the ground-state wave function $|\varPhi\rangle$ is defined as the quasi-particle vacuum $\alpha_k|\varPhi\rangle=0$, in which the quasi-particle operators $(\alpha,\alpha^\dagger)$ are connected to the original particle ones via a linear Bogoliubov transformation :
\begin{equation}
 \alpha_k=\sum_n (U_{nk}^* c_n + V_{nk}^* c_n^\dagger),~~~~~~~~~~~~\alpha_k^\dagger=\sum_n (V_{nk} c_n + U_{nk} c_n^\dagger),
\end{equation}
In terms of the normal $\rho$ and pairing $\kappa$ one-body density matrices, defined as :
\begin{equation}
\rho_{nn'}=\langle\Phi|c_{n'}^\dagger c_{n}|\Phi\rangle=(V^{*}V^{T})_{nn'}, \hspace{0.3 in}
\kappa_{nn'}=\langle\Phi|c_{n'}c_{n}|\Phi\rangle=(V^{*}U^{T})_{nn'}\,,
\label{matrices}
\end{equation}
the expectation value of the Hamiltonian (\ref{eq1}) is expressed as an energy functional
\begin{equation}
E[\rho,\kappa]=\frac{\langle\Phi|H|\Phi\rangle}{\langle\Phi|\Phi\rangle}=
\textrm{Tr}[(e+\frac{1}{2}\Gamma)\rho]-\frac{1}{2}\textrm{Tr}[\Delta\kappa^{*}]
\label{energyfunctional}
\end{equation}
where\\
\begin{equation}
 \Gamma_{n_{1}n_{3}}=\sum_{n_{2}n{4}}\bar\upsilon_{n_{1}n_{2}n_{3}n_{4}}\rho_{n_{4}n_{2}}\,,~~~~~~~~~~         \Delta_{n_{1}n_{2}}=\frac{1}{2}\sum_{n_{3}n{4}}\bar\upsilon_{n_{1}n_{2}n_{3}n_{4}}\kappa_{n_{3}n_{4}}\,.
\end{equation}
The variation of the energy (\ref{energyfunctional}) with respect to $\rho$ and $\kappa$ leads to the HFB equations:
\begin{eqnarray}
\left(\begin{array}{cc} e+\Gamma-\lambda & \Delta \\
-\Delta^* & -(e+\Gamma)^*+\lambda
\end{array} \right)\left(\begin{array}{c} U \\ V \end{array} \right) =E\left(\begin{array}{c} U \\ V \end{array}\right), \, \label{eqhfb}
\end{eqnarray}
where $\Delta$ and $\lambda$ denote the pairing potential and Lagrange multiplier, introduced to fix the correct average particle number, respectively.\\
It should be stressed that the energy functional (\ref{energyfunctional}) contains terms that cannot be simply related to some prescribed effective interaction \cite{Bender}. In terms of Skyrme forces, the HFB energy (\ref{energyfunctional})
has the form of local energy density functional:

\begin{equation}
E[\rho,\tilde{\rho}]=\int d^{3}\textrm{H}(\textbf{r}),
\label{skyrmeefunctional}
\end{equation}
where
\begin{equation}
\textrm{H}(\textbf{r})=H(\textbf{r})+\tilde{H}(\textbf{r})
\end{equation}
is the sum of the mean field and pairing energy densities. The variation of
the energy (\ref{skyrmeefunctional}) according to the particle local
density $\rho$ and pairing local density $\tilde{\rho}$ results in
Skyrme HFB equations:
\begin{eqnarray}
\sum_{\sigma'}\left(\begin{array}{cc} h(\textbf{r},\sigma,\sigma') & \tilde{h}(\textbf{r},\sigma,\sigma') \\
\tilde{h}(\textbf{r},\sigma,\sigma') & -h(\textbf{r},\sigma,\sigma')
\end{array} \right)\left(\begin{array}{c} U(E,\textbf{r}\sigma') \\
V(E,\textbf{r}\sigma')\end{array} \right) =\left(\begin{array}{cc} E+\lambda & 0\\
0 & E-\lambda)\end{array}\right)\left(\begin{array}{c} U(E,\textbf{r}\sigma) \\
V(E,\textbf{r}\sigma) \end{array}\right), \,
\label{shfb}
\end{eqnarray}
where $\lambda$ is the chemical potential. The local fields $h(\textbf{r},\sigma,\sigma')$
and $\tilde{h}(\textbf{r},\sigma,\sigma')$ can be calculated in coordinate space. Details can be found in Refs.~\refcite{Stoitsov,Ring,Greiner}.

\section{Details of Calculations}
In the present study, a parametric form of total HFB energy with Skyrme force SLy5 \cite{Chabanat} has been used as in Ref.~\refcite{Stoitsov}. Ground state properties of even-even and odd $^{124-161}Nd$ have been reproduced by using the code HFBTHO (v2.00d) \cite{Stoitsov2013} which utilizes the axial Transformed Harmonic Oscillator (THO) single-particle basis to expand quasi-particle wave functions. It iteratively diagonalizes the Hartree-Fock-Bogoliubov Hamiltonian based on generalized Skyrme-like energy densities and zero-range pairing interactions until a self-consistent solution is found. 

Calculations were performed with the SLy5 Skyrme functional, a mixed surface-volume pairing with identical pairing strength for both protons and neutrons, and a quasi-particle cutoff of $E_{cut}=60~Mev$. The Harmonic Oscillator basis was Characterized by the oscillator length $b_0=-1.0$ which means that the code automatically sets $b_0$ by using $\hbar \omega_0=1.2*41/A^{1/3}$. 
The number of oscillator shells taken into account was $N_{max}=16~shells$, the total number of states in the basis $N_{states}=500$, and the value of the deformation $\beta$ is taken from the column $\beta_2$ of the Ref.~\refcite{Moller95}. The number of Gauss-Laguerre and Gauss-Hermite quadrature points was $N_{GL} = N_{GH} = 40$, and the number of Gauss-Legendre points for the integration of the Coulomb potential was $N_{Leg} = 80$.

In the case of odd isotopes, calculations are made by using the blocking of quasi-particle states. The identification of the blocking candidate is done using the same technique as in HFODD \cite{Dobaczewski2009} : the mean-field Hamiltonian $h$ is diagonalized at each iteration and provides a set of equivalent single-particle states. Based on the Nilsson quantum numbers of the requested blocked level provided in the input file, the code identifies the index of the quasi-particle (q.p.) to be blocked by looking at the overlap between the q.p. wave-function (both lower and upper component separately) and the s.p. wave-function. The maximum overlap specifies the index of the blocked q.p.\cite{Stoitsov2013}.

There are different parameters sets of Skyrme forces for prediction of the nuclear ground-state properties \cite{Bartel,Baran}. SLy5 \cite{Chabanat} parameters set used in this study is given in Table \ref{table1}.

\begin{table}[ht]
\tbl{SLy5 parameters set.\label{table1}}
{\begin{tabular}{@{}c@{\hspace{18pt}}@{\hspace{18pt}}c@{}} \toprule
Parameter & SLy5 \\ \colrule
t$_0$ (MeV fm$^3$)   &     -2484.88 \\
t$_1$ (MeV fm$^5$)   &       462.18 \\
t$_2$ (MeV fm$^5$)   &      -448.61 \\
t$_3$ (MeV fm$^4$)   &       13673 \\
x$_0$ 		    &       0.825 \\
x$_1$    	    &       -0.465 \\
x$_2$   	    &      -1.0 \\
x$_3$  		    &       1.355 \\
W$_0$ (MeV fm$^3$)   &       126	\\
$\sigma$            &        1/6    \\  \botrule
\end{tabular}}
\end{table}

In the input data file of HFBTHO program (v2.00d) \cite{Stoitsov2013}, we have modified the values of the pairing strength for neutrons $V_0^{n}$ and protons $V_0^{p}$ (in MeV), which may be different, but in our study we have used the same pairing strength $V_0^{n,p}$ for both. At each time, we have executed the program and compared the obtained ground-state energy with the experimental value. This procedure was repeated until we found the value of $V_0^{n,p}$ that gives the ground-state energy closest to the experimental one.

The calculated ground-state energies of $^{124-161}Nd$ isotopes, obtained in this
work with the corresponding pairing strength $V_0^{n,p}$, and the experimental data\cite{WANG} are listed in Table~\ref{table2}.

\begin{table}[!ht]
\centering
\tbl{The ground-state energies of $^{124-161}Nd$ isotopes (in units of MeV) obtained in this work by using HFB method with SLy5 Skyrme force.\label{table2}}
{\begin{tabular}{@{}c@{\hspace{10pt}}c@{\hspace{10pt}}c@{\hspace{10pt}}c@{}} \toprule
Nuclei      & Experiment  	&	Calculat   & $V_0^{n,p}$\\ \colrule
$^{124}$Nd  &   998.448         &   998.4436       &    380.9   \\
$^{125}$Nd  &   1009.625        &   1009.6251       &    389.8   \\
$^{126}$Nd  &   1022.994        &   1022.9924       &    381.8   \\
$^{127}$Nd  &   1033.653        &   1033.6392       &    389.4   \\
$^{128}$Nd  &   1046.528        &   1046.5205       &    384.0   \\
$^{129}$Nd  &   1056.51         &   1056.5037       &    392.4   \\
$^{130}$Nd  &   1068.9263       &   1068.9317       &    386.3   \\
$^{131}$Nd  &   1078.1693       &   1078.1600       &    390.6   \\
$^{132}$Nd  &   1089.8989       &   1089.8940       &    384.1   \\
$^{133}$Nd  &   1098.8726       &   1098.8726       &    392.3   \\
$^{134}$Nd  &   1110.2623       &   1110.2661       &    382.0   \\
$^{135}$Nd  &   1118.9002       &   1118.9133       &    387.1   \\
$^{136}$Nd  &   1129.9573       &   1129.9546       &    380.8   \\
$^{137}$Nd  &   1138.4138       &   1138.4186       &    388.2   \\
$^{138}$Nd  &   1148.919        &   1148.9115       &    379.9   \\
$^{139}$Nd  &   1156.9873       &   1156.9768       &    386.1   \\
$^{140}$Nd  &   1167.2976       &   1167.2870       &    370.4   \\
$^{141}$Nd  &   1175.3083       &   1175.3016       &    391.5   \\
$^{142}$Nd  &   1185.1361       &   1185.1350       &    377.7   \\  \botrule
\end{tabular}
\quad \quad
\begin{tabular}{@{}c@{\hspace{10pt}}c@{\hspace{10pt}}c@{\hspace{10pt}}c@{}} \toprule
Nuclei      & Experiment  	&	Calculat   & $V_0^{n,p}$\\ \colrule
$^{143}$Nd  &   1191.2596        &   1191.2528       &    393.1   \\
$^{144}$Nd  &   1199.0767        &   1199.0689       &    376.7   \\
$^{145}$Nd  &   1204.8319        &   1204.8252       &    392.9   \\
$^{146}$Nd  &   1212.3972        &   1212.3944       &    386.4   \\
$^{147}$Nd  &   1217.6894        &   1217.6898       &    392.9   \\
$^{148}$Nd  &   1225.0219        &   1225.0129       &    385.6   \\
$^{149}$Nd  &   1230.0607        &   1230.0599       &    392.2   \\
$^{150}$Nd  &   1237.4358        &   1237.4304       &    387.7   \\
$^{151}$Nd  &   1242.7704        &   1242.7757       &    398.0   \\
$^{152}$Nd  &   1250.048         &   1250.0499       &    392.9   \\
$^{153}$Nd  &   1255.301         &   1255.3112       &    403.9   \\
$^{154}$Nd  &   1261.722         &   1261.7247       &    396.1   \\
$^{155}$Nd  &   1266.3965        &   1266.3885       &    405.4   \\
$^{156}$Nd  &   1272.6636        &   1272.6725       &    398.8   \\
$^{157}$Nd  &   1276.7177        &   1276.7125       &    406.8   \\
$^{158}$Nd  &   1282.328         &   1282.3388       &    399.1   \\
$^{159}$Nd  &   1286.151         &   1286.1507       &    406.7   \\
$^{160}$Nd  &   1291.68          &   1291.6796       &    400.3   \\
$^{161}$Nd  &   1295.084         &   1295.0885       &    408.0   \\  \botrule
\end{tabular}
}
\end{table}

As can be noted from Table \ref{table2}, there is a relationship between the pairing strength $V_0^{n,p}$ and the mass number $A$. By fitting the obtained values of $V_0^{n,p}$ to  $A$, we have found the following formula :
 \begin{equation}
 \boxed{
\large {{ V_0^{n,p} = 170.95\,A^{\frac{1}{6}}}}
 }
 \label{eqV0}
 \end{equation}

\begin{figure}[th]
\centerline{\psfig{file=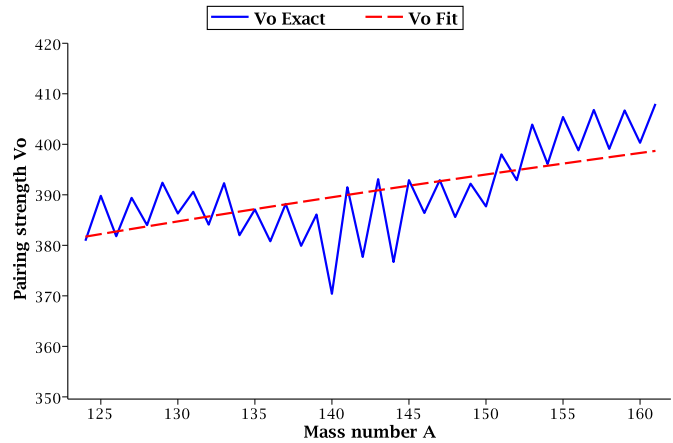,width=9cm}}
\caption{The exact and adjusted pairing-strength $V_0^{n,p}$ .}
 \label{pairing}
\end{figure}

On Fig.\ref{pairing}, we present two curves which show the variation of $V_0$ as a
 function of the mass number $A$. The solid curve is obtained from the data of Table \ref{table2}, and the dashed one is the graphical representation of Eq.~(\ref{eqV0}). The mean deviation between $V_{0_{fit}}$ and $V_{0_{exact}}$ is about $5.05$ Mev.

In order to verify the validity of Eq.~(\ref{eqV0}), we have used this equation to generate the pairing-strength $V_0^{n,p}$ that we have included in the code HFBTHO (v2.00d) in order to calculate the ground-state properties for both even-even and odd $^{124-161}Nd$
isotopes. Also, the same calculations have been performed for $^{128-165}Sm$ and $^{119-157}Ce$ isotopes. The results are presented in the next section.

\section{Results and Discussion}
In this section we present the numerical results of this work, particularly for binding energy, two-neutron and two-proton separation energies and charge and neutron radii for $^{124-161}Nd$,$^{128-165}Sm$ and $^{119-157}Ce$ isotopes.\\
In all our calculations, we used the Skyrme (SLy5) force and Eq.(\ref{eqV0}) for the pairing strength.

\subsection{Binding energy}

\begin{figure}[ht]
\centerline{\psfig{file=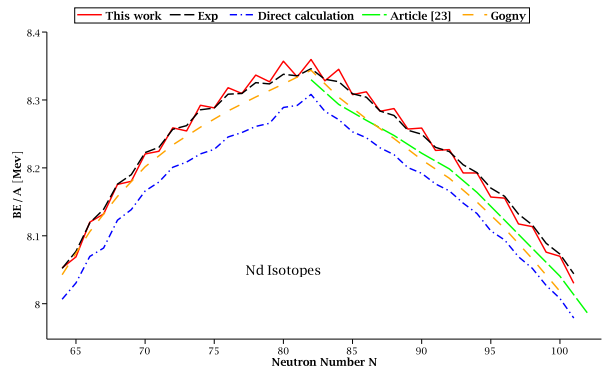,width=8cm}}
\caption{Binding energies per nucleon for even-even and odd isotopic chains of $Nd$ nuclei.}
 \label{BEexp}
\end{figure}

In Fig.\ref{BEexp}, the calculated Binding Energy (BE) per nucleon for $Nd$ isotopes, obtained by using the pairing strength generated by Eq.~(\ref{eqV0}) as well as by direct calculations using the pre-defined pairing strength in  HFBTHO(v2.00d) program \cite{Stoitsov2013} are shown.
Also, in Fig.\ref{BEexp}, we present the experimental binding energies per nucleon for $Nd$ isotopes \cite{WANG}, the results of HFB calculations based on the D1S Gogny force\cite{AMEDEE} and the obtained results in Ref.~\refcite{Article} in which the authors have used the code HFBTHO (v1.66p) \cite{Stoitsov} to reproduce ground-state properties of even-even $^{142-164}Nd$ isotopes.

From Fig.\ref{BEexp} we note that the maximum in the BE per nuclei is observed at the magic neutron number $N = 82$ in experimental data as well as in the HFB calculations based on the D1S Gogny force\cite{AMEDEE} and HFB method with SLy5 Skyrme force for both direct calculations and calculations with Eq.~(\ref{eqV0}). It should also be noted  that using Eq.~(\ref{eqV0}) gives improved results of the binding energy, and therefore, it will ameliorate the results of other ground-state properties of the nuclei.

\begin{figure}[ht]
\centerline{\psfig{file=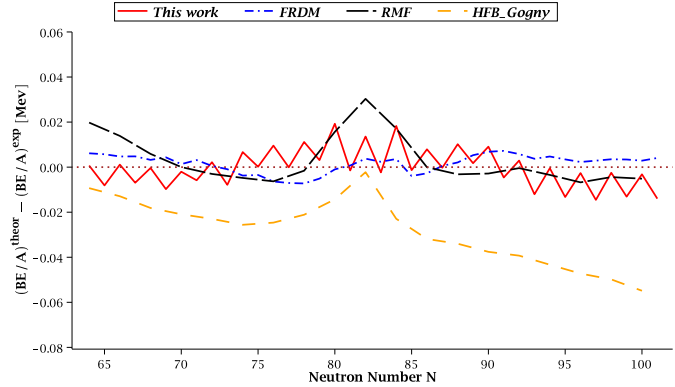,width=9cm}}
\caption{The differences between our calculated results for binding energies per nucleon within HFB theory and experimental values \cite{WANG}, HFB calculations based on the D1S Gogny force\cite{AMEDEE}  and the predictions of both FRDM \cite{Moller97} and the RMF \cite{Lalazissis} theories are shown for comparison.}
 \label{Diff_BE}
\end{figure}

The differences between the experimental BE per nucleon and the calculated results obtained in this work by using Eq.~(\ref{eqV0}) are shown as function of the neutron number $N$ in Fig.\ref{Diff_BE}. The HFB calculations based on the D1S Gogny force \cite{AMEDEE} as well as the predictions of Finite Range Droplet Model (FRDM) \cite{Moller97} and Relativistic Mean Field (RMF) model with NL3 functional \cite{Lalazissis} are also included for comparison.

As can be seen in Fig.\ref{Diff_BE}, the calculated BE per nucleon for $Nd$ isotopes are in a good agreement with the experimental data. The maximal error is about $0.019$ $MeV$ per particle which corresponds approximately to $2.697$ $MeV$ for the total binding energy.

\begin{figure}[!htb]
\minipage{0.48\textwidth}
\centerline{\psfig{file=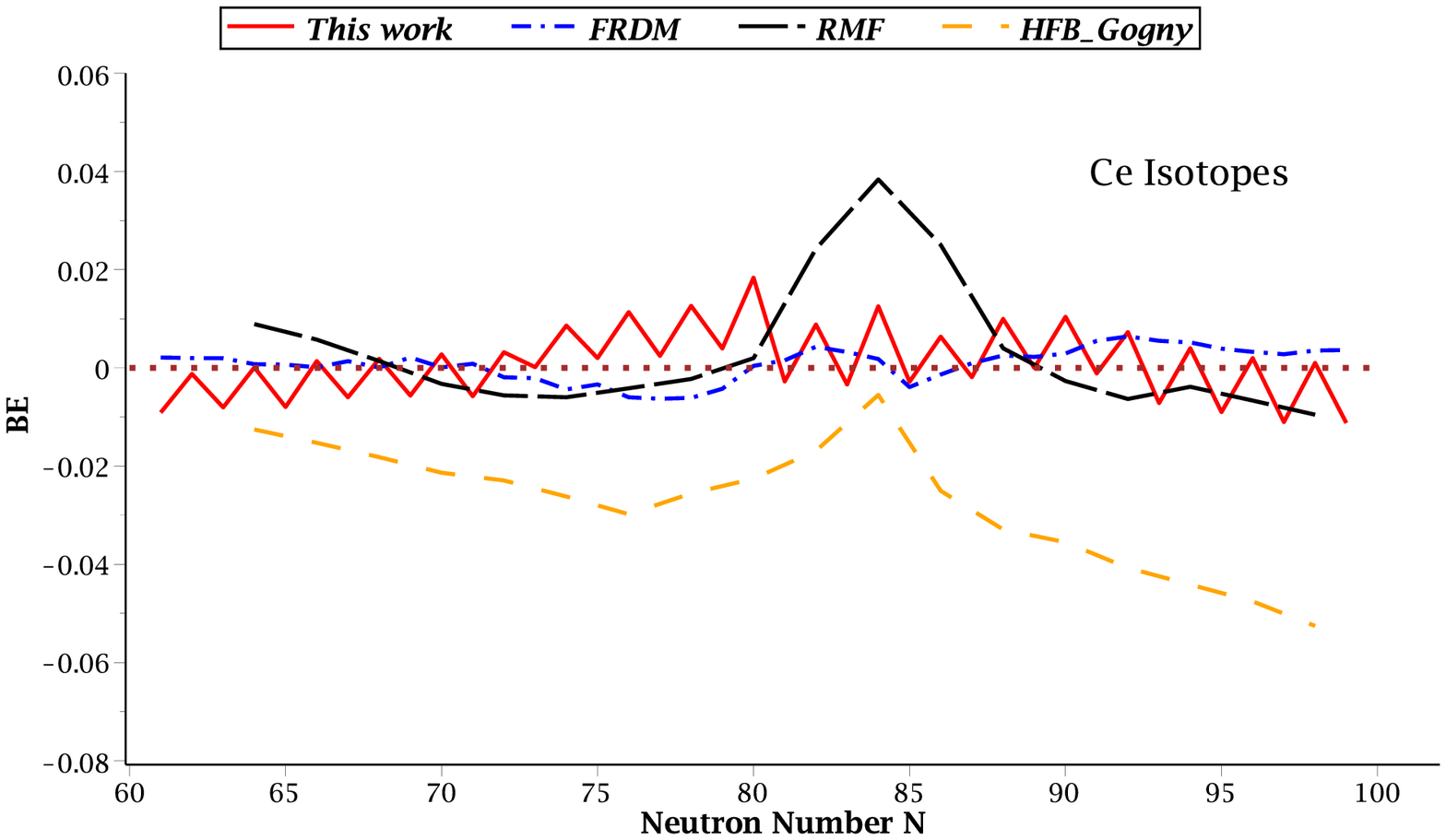,width=\linewidth, height=4cm}}
\endminipage\hfill
\minipage{0.48\textwidth}
\centerline{\psfig{file=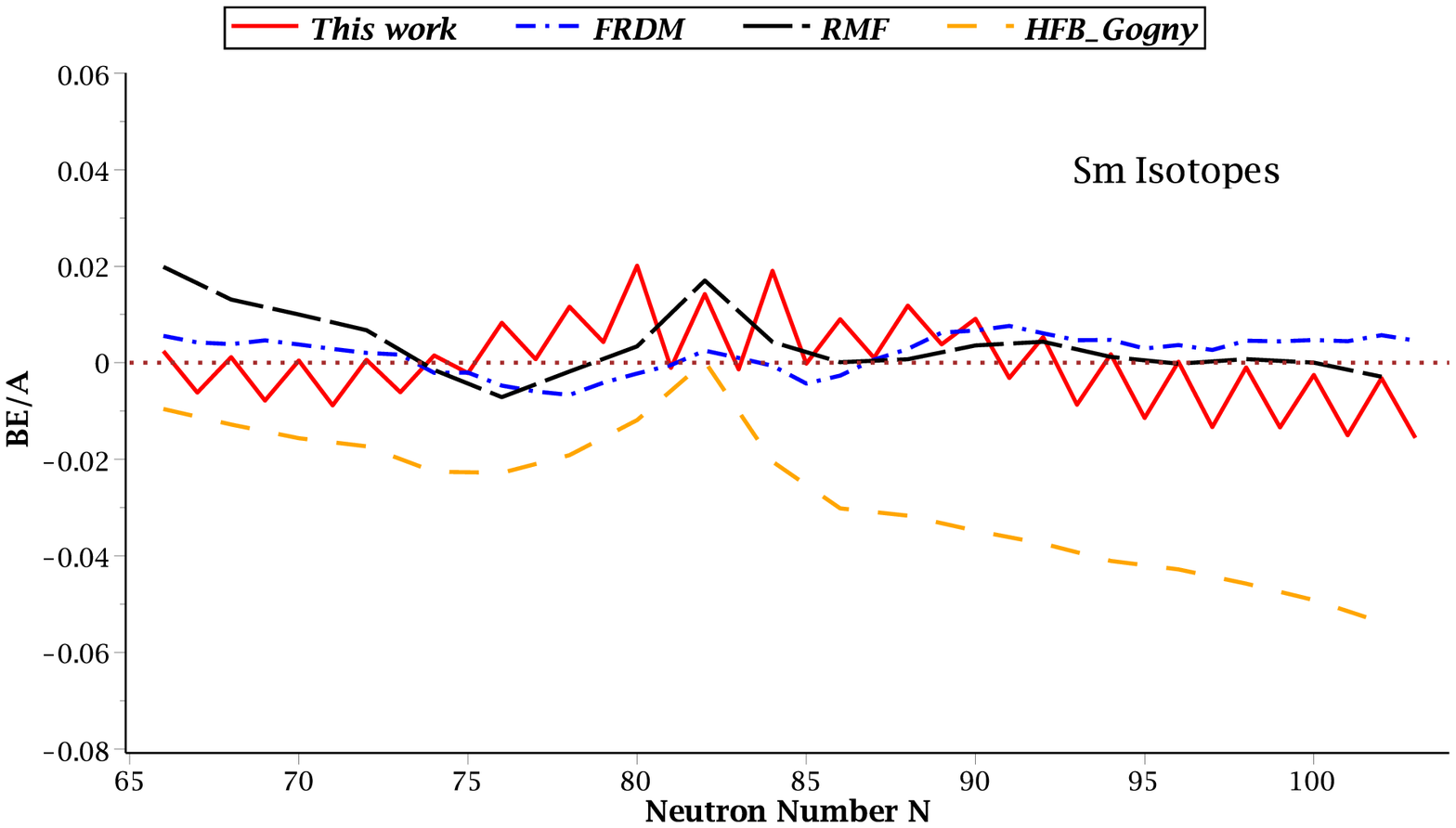,width=\linewidth, height=4cm}}
\endminipage\hfill
\caption{Same as Fig.\ref{Diff_BE}, but for $Ce$ isotopes (left) and $Sm$ isotopes (right).}
\label{BE2}
\end{figure}

In order to ensure the validity of Eq.~(\ref{eqV0}), we have used this equation to generate the pairing strength for neutrons and protons to calculate the ground state properties of two isotopic chains $^{119-157}Ce$ and $^{128-165}Sm$ which are in the vicinity of $Nd$.
The differences between our calculated results for binding energies per nucleon within HFB theory and experimental values are displayed in Fig.\ref{BE2}. The results of HFB calculations based on the D1S Gogny force\cite{AMEDEE} as well as the predictions of the FRDM \cite{Moller97} and the RMF \cite{Lalazissis} theories are displayed for comparison.

As can be seen in Fig.\ref{BE2}, the calculated BE per nucleon for $Ce$ and $Sm$ isotopes in the HFB theory with SLy5 Skyrme force are in a good agreement with the experimental data in comparison with the other theoretical predictions.
The approximately maximal errors per particle are $0.018 ~MeV$ and $0.020 ~MeV$ for $Ce$ and $Sm$ isotopes, respectively.

\begin{figure}[!ht]
\centerline{\psfig{file=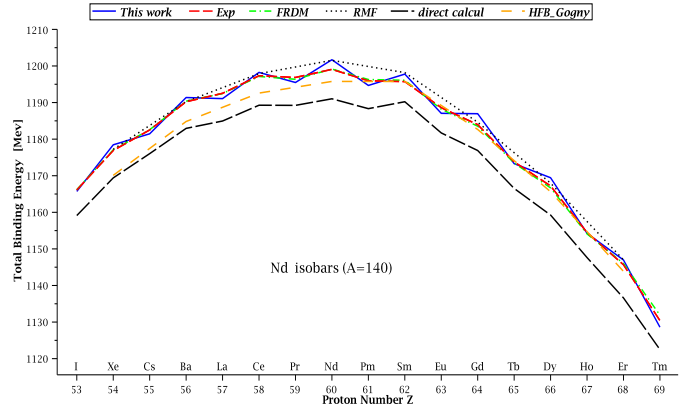,width=9cm, height=5.5cm}}
\caption{The total binding energies of $^{140}$Nd isobars}
\label{isobare_Nd}
\end{figure}

As a further verification of the validity of Eq.~(\ref{eqV0}), we have used it to generate the pairing strength for the mass number $A=140$ in order to calculate the total binding energy for even-even and odd isobars of $^{140}\mathrm{Nd}$, from $^{140}_{~53}\mathrm{I}_{91}$ to $^{140}_{~69}\mathrm{Tm}_{75}$.
 The calculated total binding energies and experimental data \cite{WANG} for this series of isobars are given as function of the proton number $Z$ in Fig.\ref{isobare_Nd}. The direct calculations as well as the HFB calculations based on the D1S Gogny force\cite{AMEDEE} and the predictions of both RMF \cite{Lalazissis} and FRDM \cite{Moller97} theories are shown for comparison.

From Fig.\ref{isobare_Nd} we note that our calculated results for total binding energies of $^{140}$Nd isobars are in a good agreement with those from experiment and FRDM and RMF theories. The use of Eq.~(\ref{eqV0}) has improved the results of total binding energies compared to those of direct calculations. The mean absolute error between experimental data and results of this work is $1.436$ Mev, while it is $0.347$ Mev, $1.076$ Mev, $3.123$ and $7.392$ Mev in FRDM\cite{Moller97}, RMF\cite{Lalazissis} theory, HFB calculations based on the D1S Gogny force\cite{AMEDEE} and direct calculations, respectively.

\subsection{Neutron separation energy}
The one-neutron and two-neutron separation energies are important quantities to exhibit the nuclear shell structure. In the present work, we calculated two-neutron separation energies ($S_{2n}$) for $Nd$, $Ce$ and $Sm$ isotopes in SLy5 parametrization with the pairing strength $V_0^{n,p}$ generated by Eq.~(\ref{eqV0}).

The two-neutron separation energy is defined as
\begin{equation}
 S_{2n}(Z,N)=BE(Z,N)-BE(Z,N-2)
\end{equation}
Note that when using this equation, all binding energies must be involved with a positive sign.

\begin{figure}[!htb]
\minipage{0.33\textwidth}
\centerline{\psfig{file=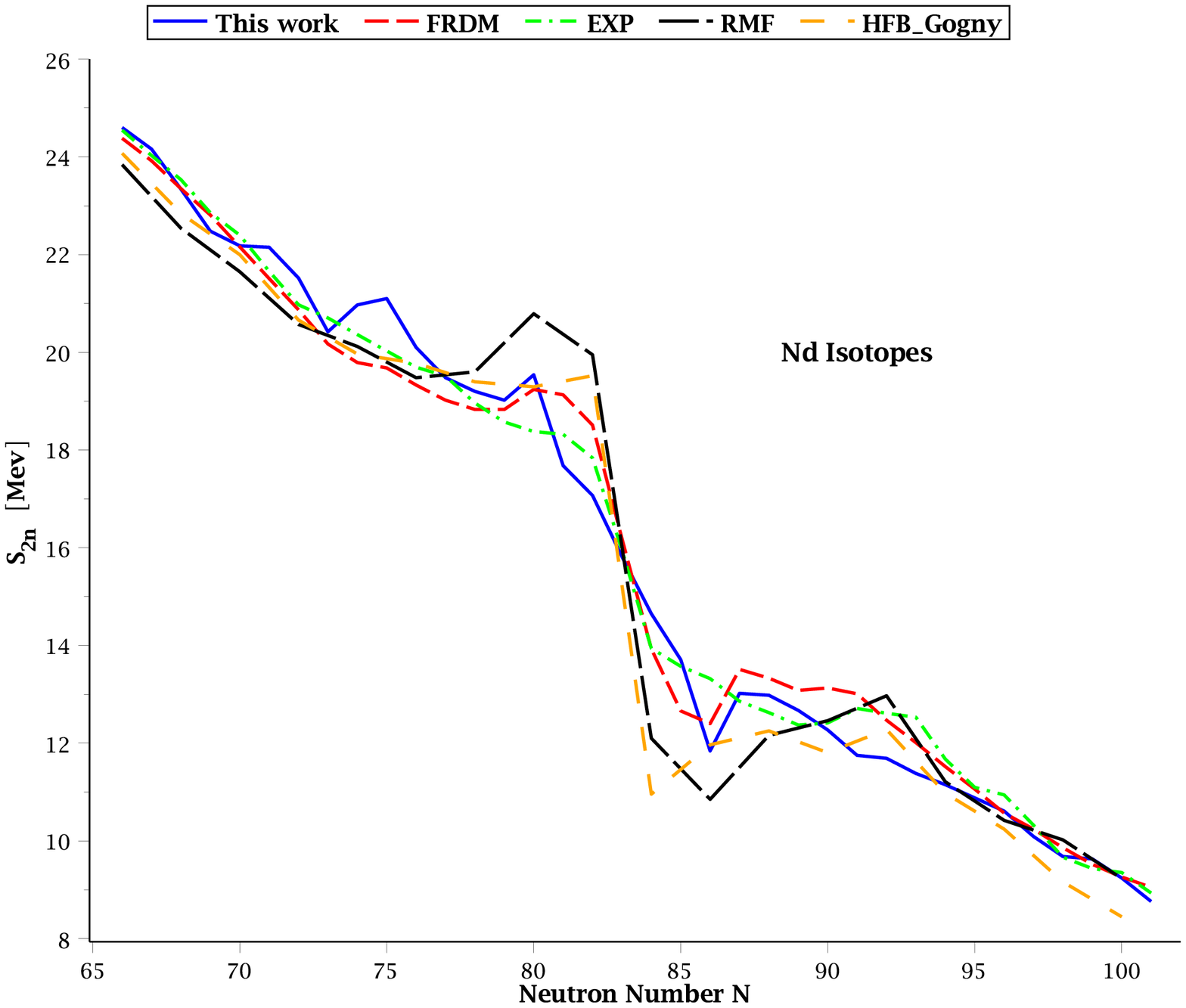,width=4cm, height=4.5cm}}
\endminipage\hfill
\minipage{0.33\textwidth}
\centerline{\psfig{file=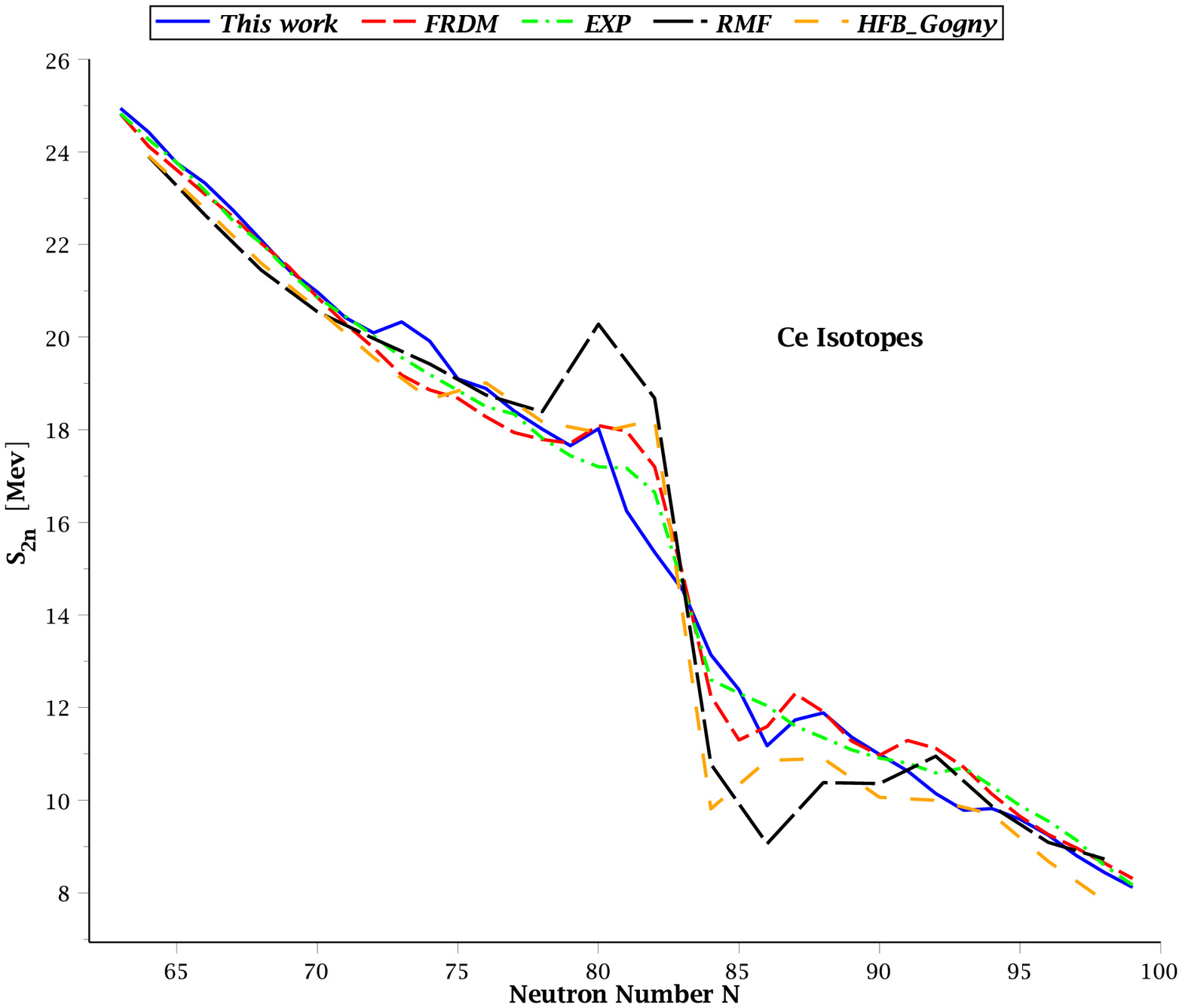,width=4cm, height=4.5cm}}
\endminipage\hfill
\minipage{0.33\textwidth}%
\centerline{\psfig{file=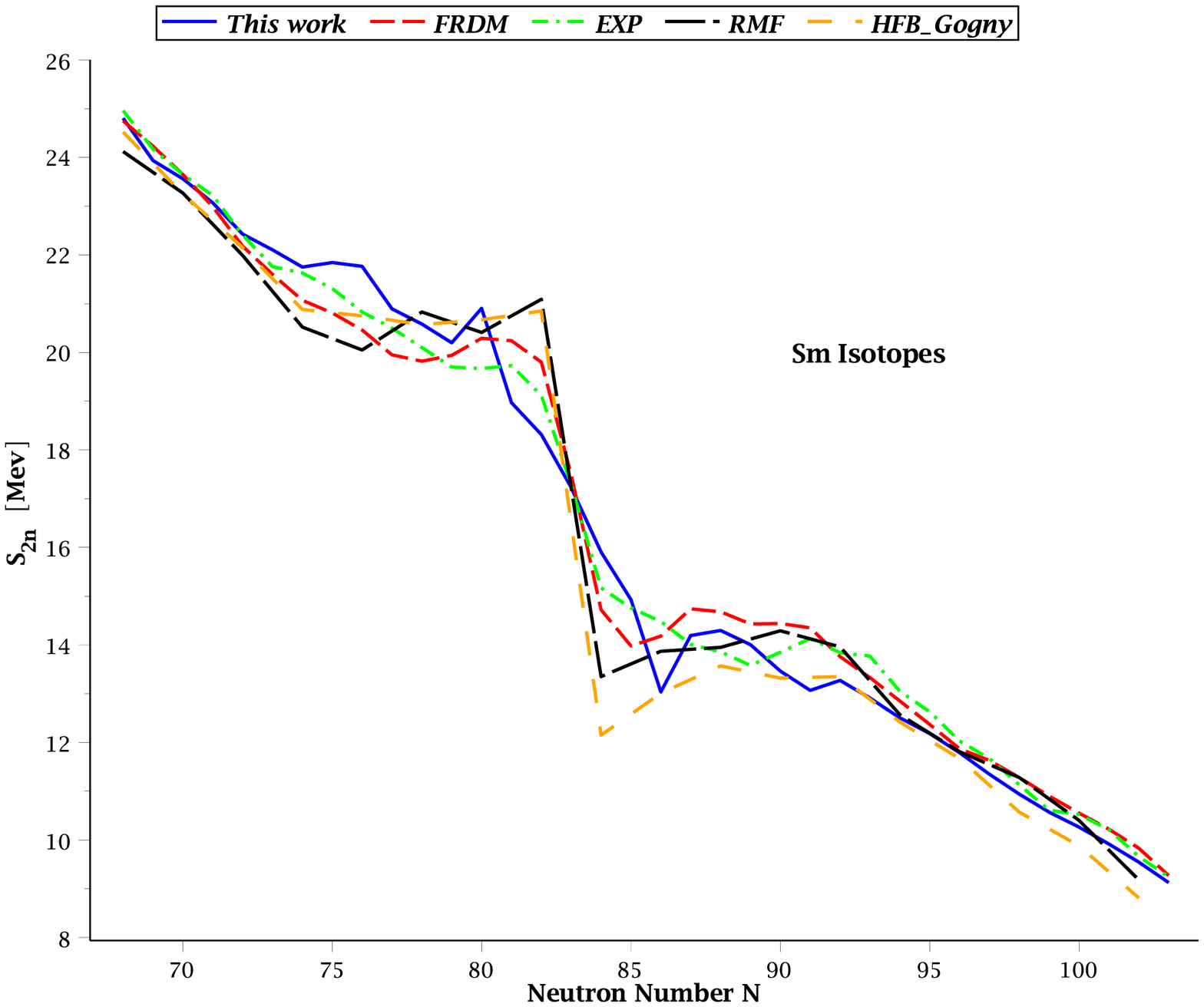,width=4cm, height=4.5cm}}
\endminipage
\caption{Comparison of the calculated two-neutron separation energies $S_{2n}$ of $Nd$, $Ce$ and $Sm$ isotopes with the RMF\cite{Lalazissis} model, FRDM\cite{Moller97}, experimental data\cite{WANG} and HFB calculations based on the D1S Gogny force\cite{AMEDEE}.}
\label{S2n}
\end{figure}

The calculated $S_{2n}$ for $Nd$, $Ce$ and $Sm$ isotopes as well as experimental data\cite{WANG}, predictions of RMF\cite{Lalazissis} and FRDM\cite{Moller97} theories, and HFB calculations based on the D1S Gogny force\cite{AMEDEE} are presented in Fig.\ref{S2n}.

As can be seen in Fig.\ref{S2n}, a sharp decrease in $S_{2n}$ at the magic neutron number $N=82$ corresponds to the closed shell. Also, the calculated two-neutron separation energies for $Nd$, $Ce$ and $Sm$
nuclei in HFB method with SLy5 Skyrme force as well as predictions of RMF theory with NL3 parameters set and FRDM theory are
in a good agreement with experimental data. There are small differences between our results of HFB method with Skyrme force SLy5 and experimental data.

The approximately maximal errors of $S_{2n}$ between the calculated results in the
present study and experimental data for the three nuclei $Nd$, $Ce$ and $Sm$ are listed in Table \ref{table3}. The predictions of FRDM and RMF theories as well as the HFB calculations based on the D1S Gogny force\cite{AMEDEE} are listed too for comparison.

\begin{table}[ht]
\tbl{The maximal difference error $(S_{2n})_{theor}-(S_{2n})_{exp}$ (in Mev).\label{table3}}
{\begin{tabular}{@{}c@{\hspace{18pt}}c@{\hspace{18pt}}c@{\hspace{18pt}}c@{\hspace{18pt}}c@{}} \toprule
Nuclei 		&    This work   	&      	RMF    		&      FRDM		&	HFB$_{Gogny}$\\ \colrule
Nd		&	1.4782 		&	2.4705		&	0.9205		&	2.98460	\\
Ce		&	1.29951		&	3.07735		&	1.01289		&	2.78419	\\
Sm		&	1.44676		&	1.96853		&	0.84682		&	3.02534	\\ \botrule
\end{tabular}
}
\end{table}

\subsection{Neutron, Proton and Charge radii}

The root-mean-square (rms) charge radius, $R_c$, is related to the proton radius, $R_p$, by
\begin{equation}
 R_c^2=R_p^2+0.64~(fm)
\end{equation}

\begin{figure}[!htb]
 \minipage{0.33\textwidth}
 \centerline{\psfig{file=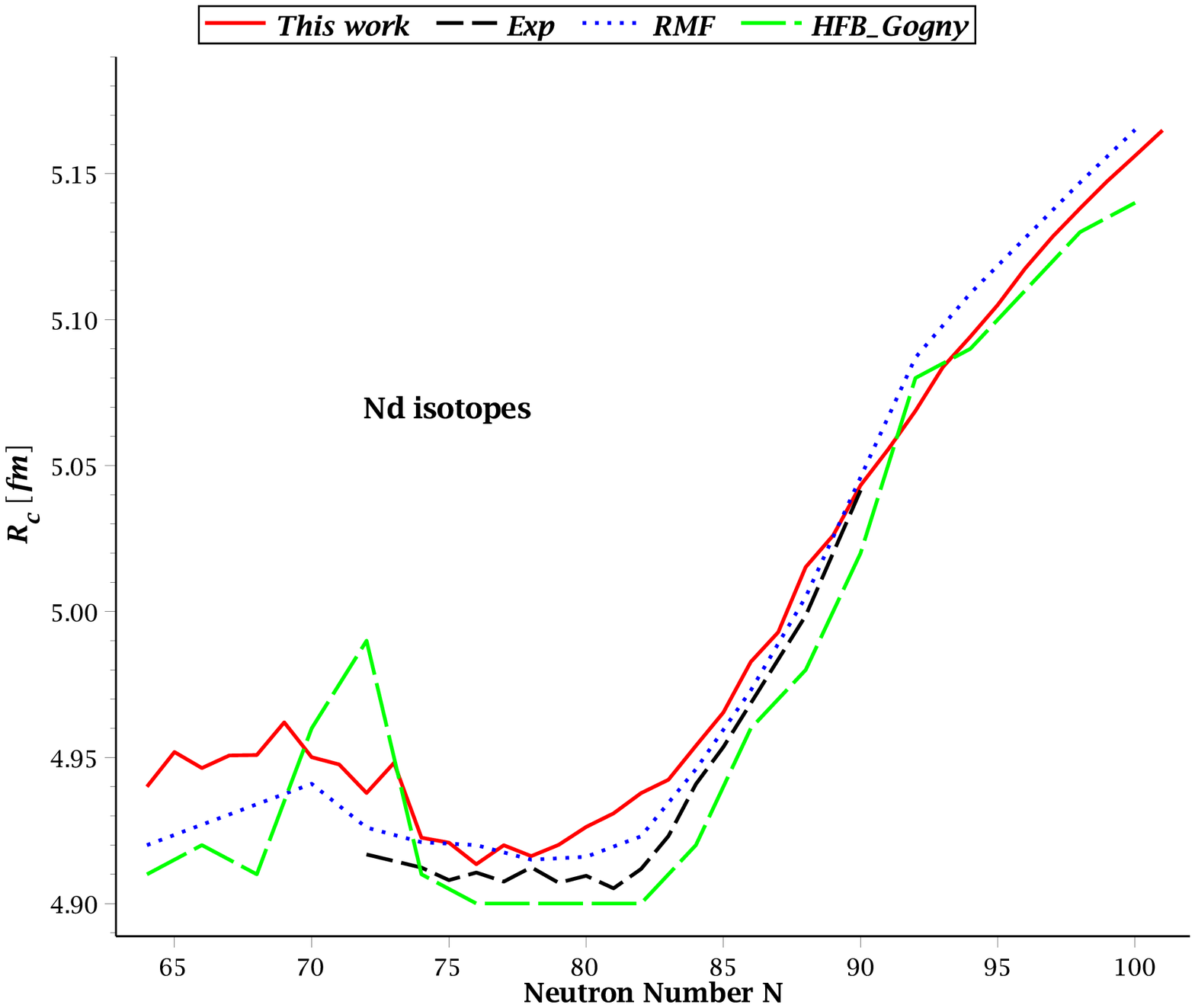,width=4cm, height=4cm}}
 \endminipage\hfill
 \minipage{0.33\textwidth}
 \centerline{\psfig{file=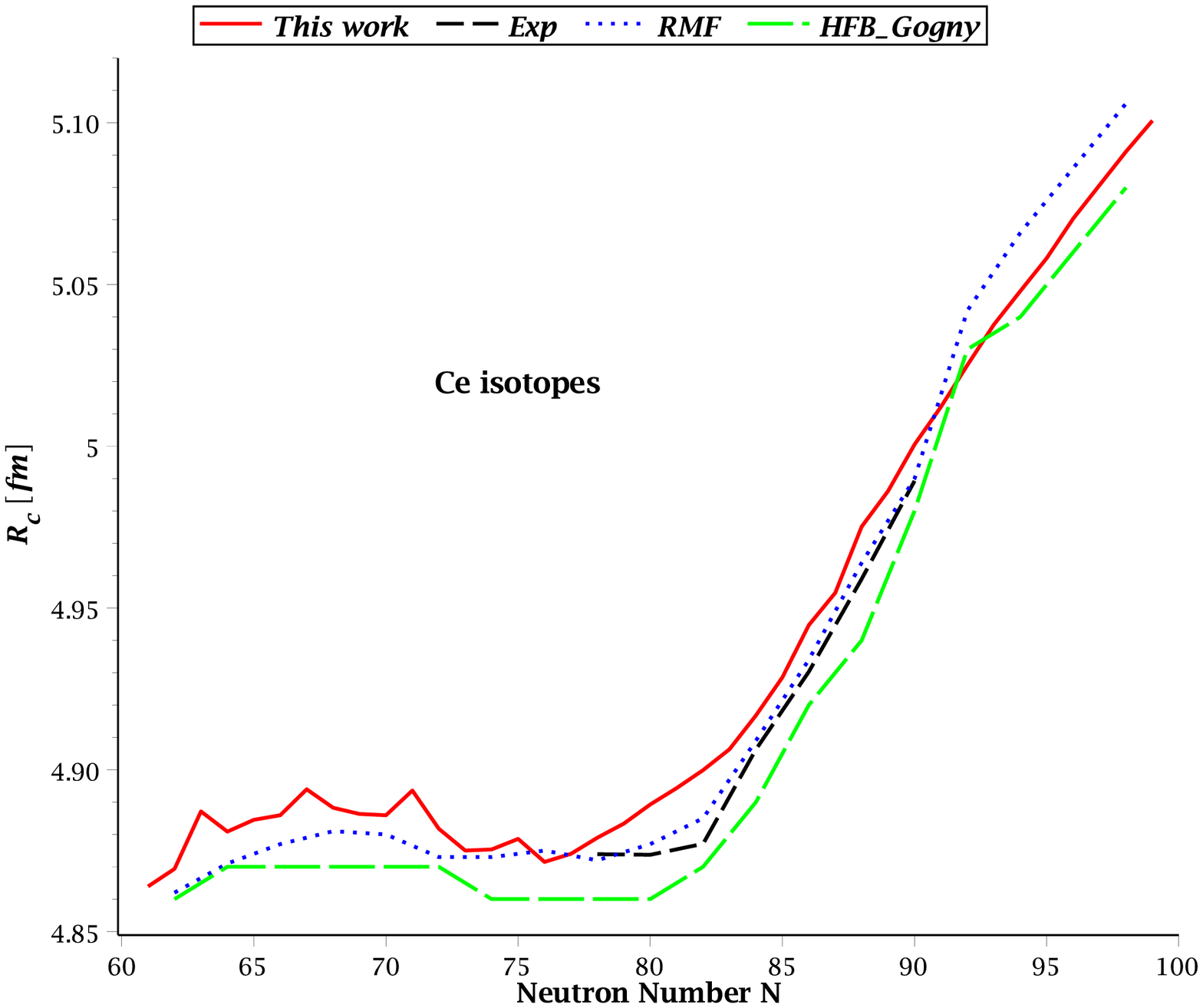,width=4cm, height=4cm}}
 \endminipage\hfill
 \minipage{0.33\textwidth}%
 \centerline{\psfig{file=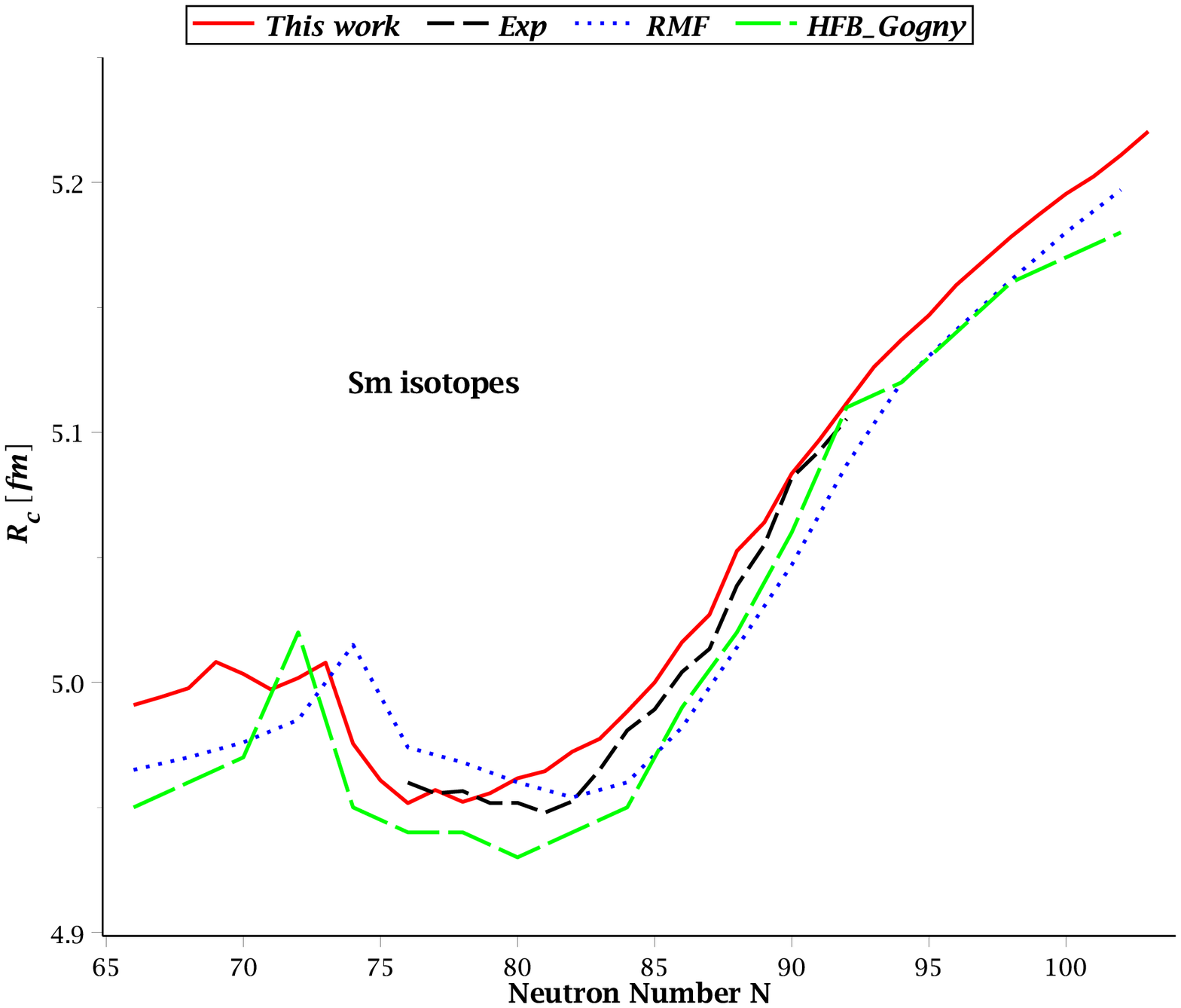,width=4cm, height=4cm}}
 \endminipage
 \caption{The charge radii obtained by our HFB calculations compared with the available experimental data \cite{Angeli}, predictions of RMF theory \cite{Lalazissis} and the HFB calculations based on the D1S Gogny force\cite{AMEDEE}.}
 \label{Rc}
 \end{figure}

In Fig.\ref{Rc} the root-mean-square charge radii predicted by our HFB calculations are compared with the available experimental data \cite{Angeli}, the predictions of RMF theory \cite{Lalazissis} and the HFB calculations based on the D1S Gogny force\cite{AMEDEE}.

A good agreement between theory and experiment can be clearly seen in Fig.\ref{Rc}.
From this figure, one can see that the charge radii slightly decrease to the magic number when going from lighter to heavier isotopes.
Therefore, the lighter isotopes have larger charge radii than the heavier closed-neutron-shell (N=82) nucleus.
The charge radii of nuclei which are heavier than the closed-neutron-shell increase as well as the neutron number increases.

\begin{figure}[!htb]
 \minipage{0.33\textwidth}
 \centerline{\psfig{file=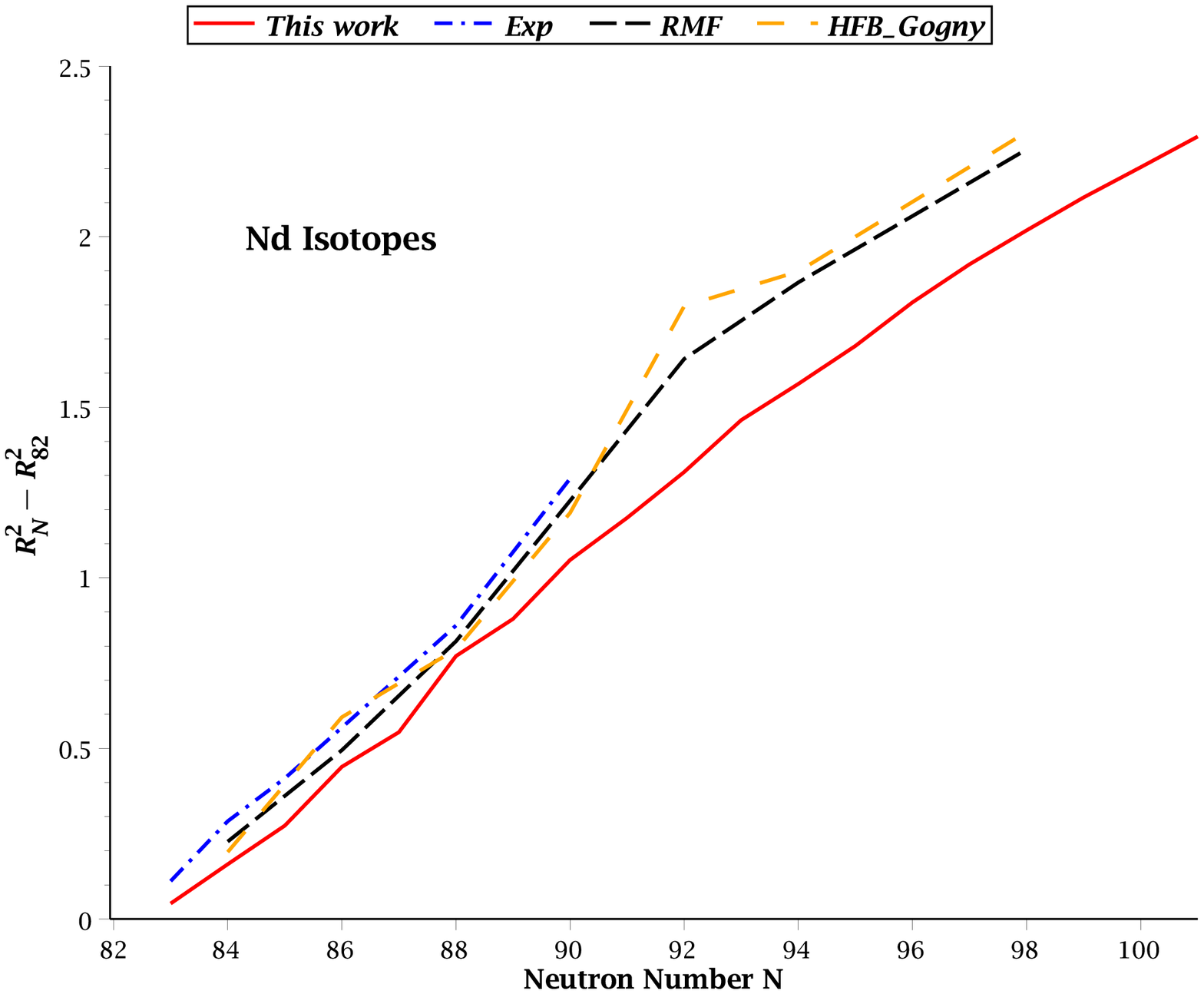,width=4cm, height=4cm}}
 \endminipage\hfill
 \minipage{0.33\textwidth}
 \centerline{\psfig{file=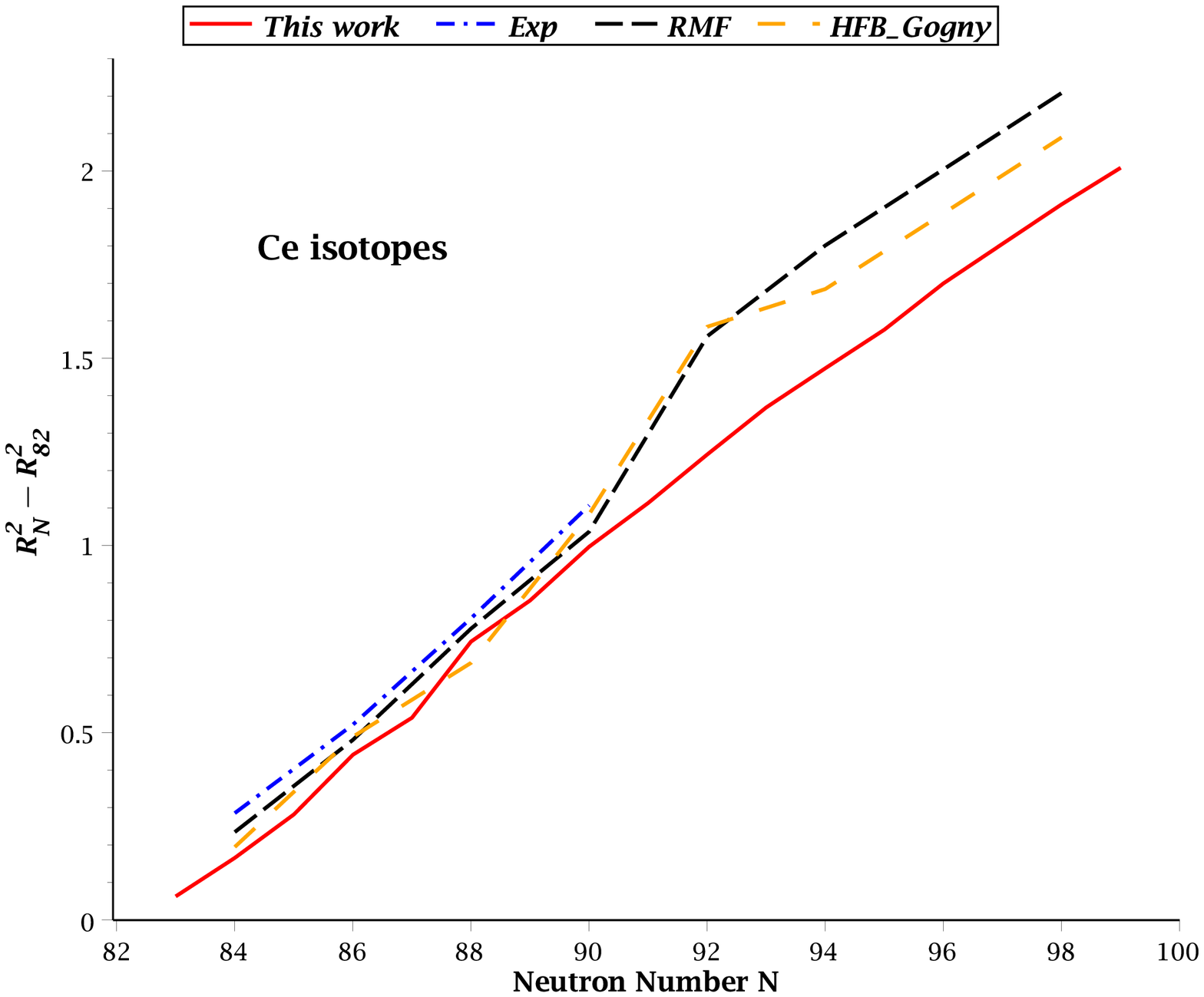,width=4cm, height=4cm}}
 \endminipage\hfill
 \minipage{0.33\textwidth}%
 \centerline{\psfig{file=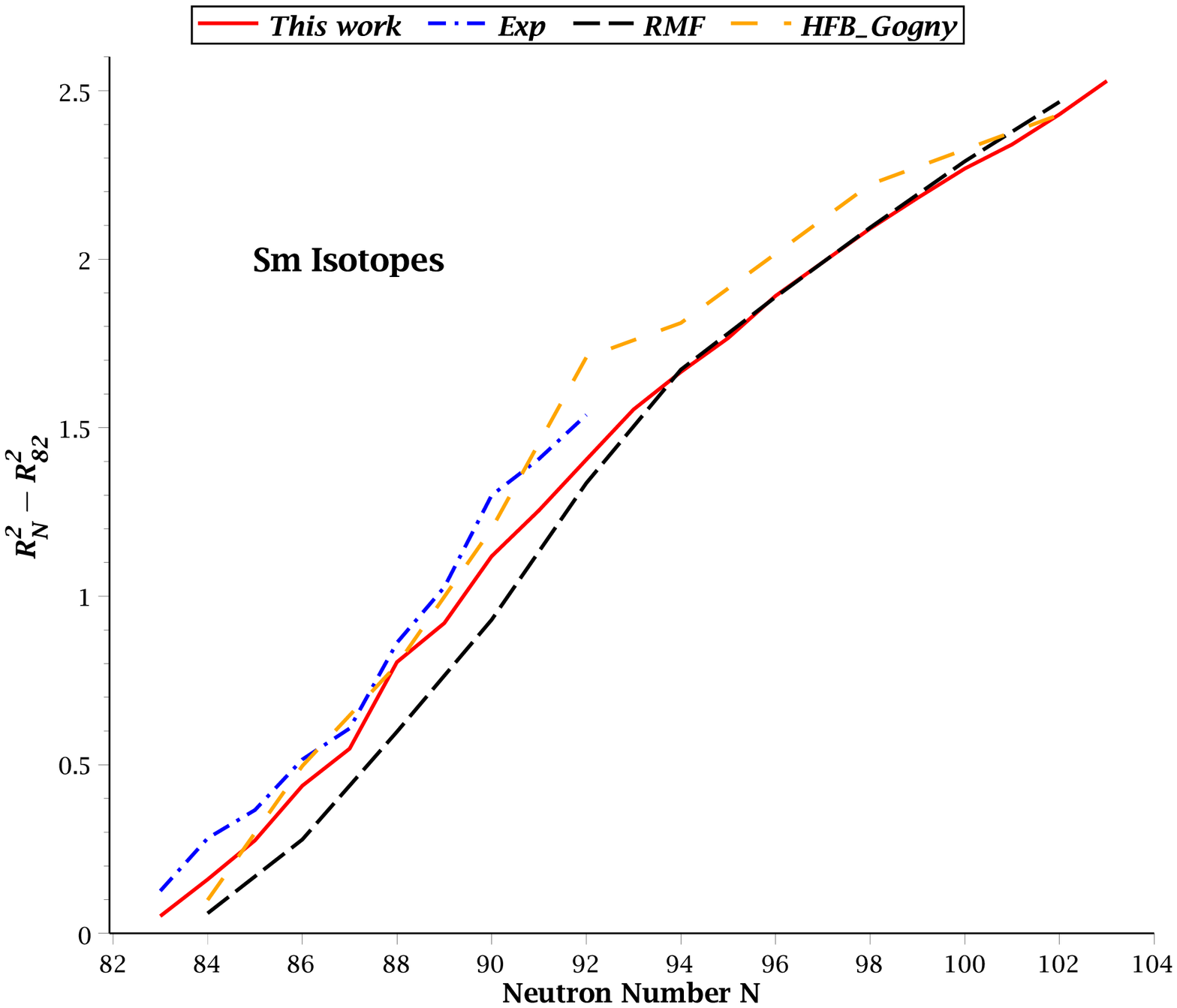,width=4cm, height=4cm}}
 \endminipage
 \caption{The differences between squares of ground state charge radii : $\langle r_N^2\rangle-\langle r_{N=82}^2\rangle$ as function of neutron number.}
 \label{R2}
 \end{figure}

In order to better understand the structural evolution of $Nd$ isotopes with increasing neutron number,  the differences between squares of ground-state charge radii of $Nd$ isotopes and those of the reference nucleus (The magic neutron number $N=82$) have been calculated. The same calculations have been performed for $^{128-165}Sm$ and $^{119-157}Ce$ isotopes. The calculated and available experimental data \cite{Angeli} of $\langle r_N^2\rangle - \langle r_{N=82}^2\rangle$ are shown in Fig.\ref{R2}. The predictions of the RMF \cite{Lalazissis} theory and the HFB calculations based on the D1S Gogny force\cite{AMEDEE} are presented for comparison.

From Fig.\ref{R2}, it is seen a good agreement between the calculated results of $\langle r_N^2\rangle-\langle r_{N=82}^2\rangle$, the predictions of RMF\cite{Lalazissis} theory, the HFB calculations based on the D1S Gogny force\cite{AMEDEE} and the available experimental data\cite{Angeli}.

\begin{figure}[!htb]
 \minipage{0.33\textwidth}
 \centerline{\psfig{file=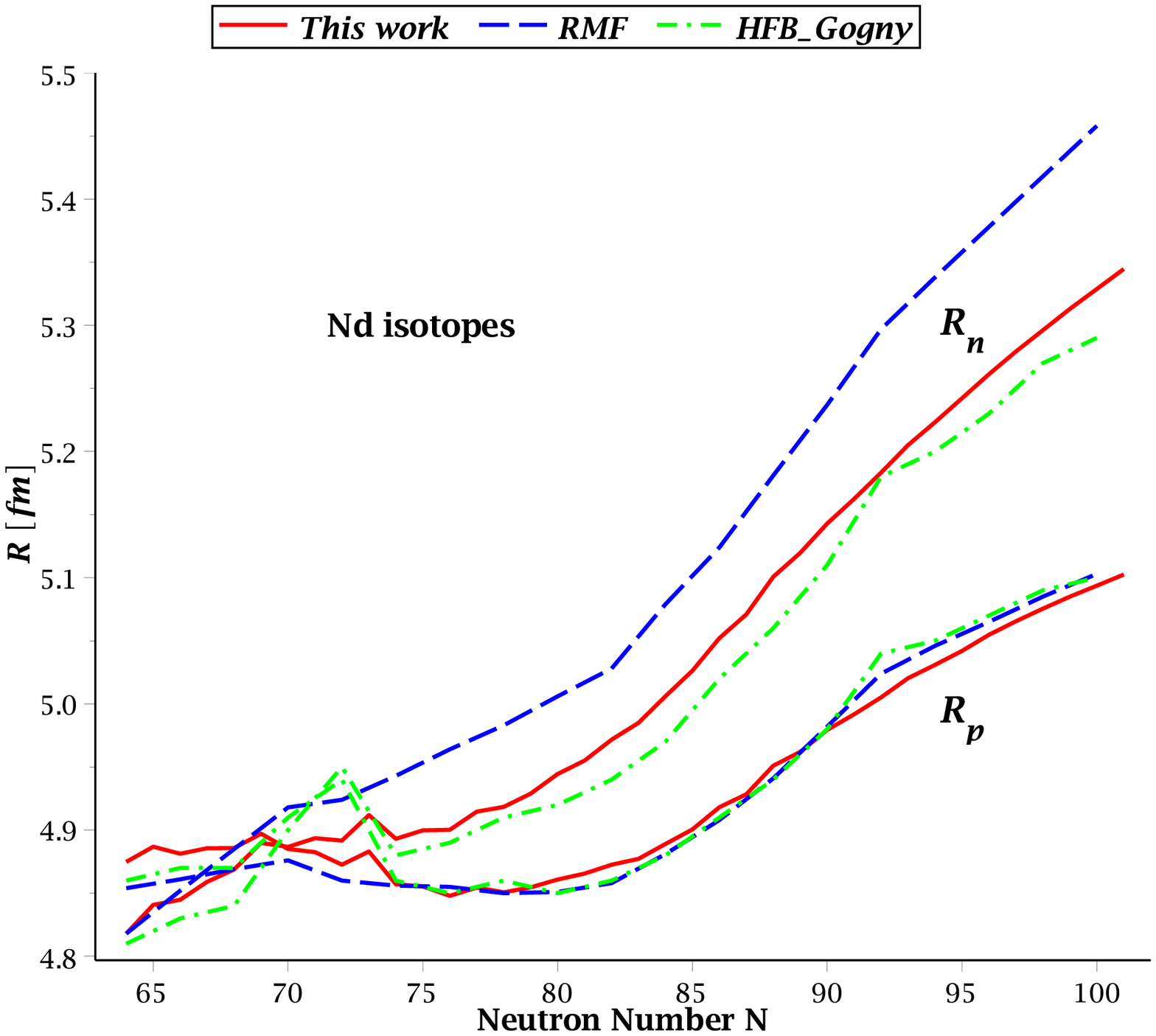,width=4cm, height=4cm}}
 \endminipage\hfill
 \minipage{0.33\textwidth}
 \centerline{\psfig{file=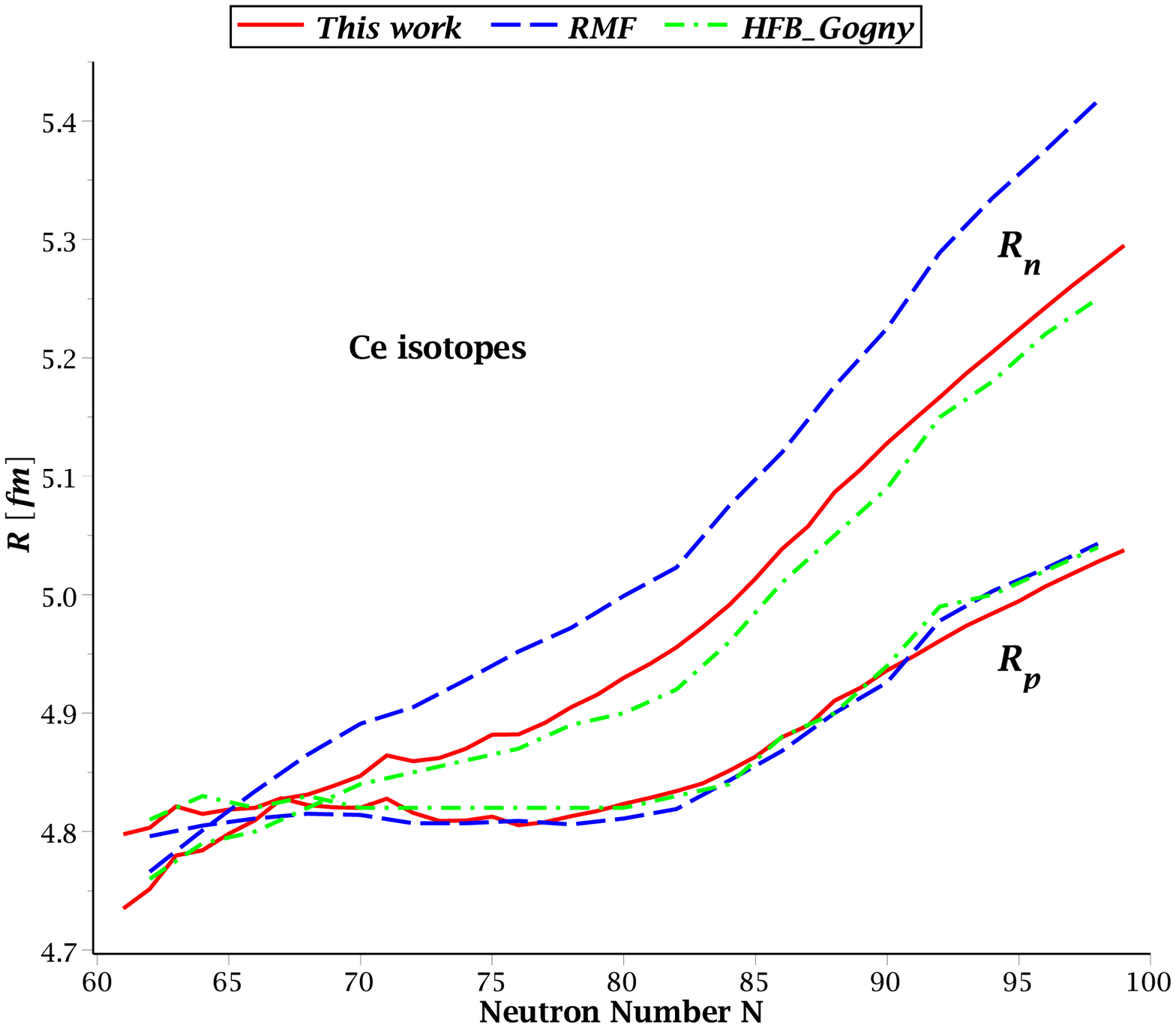,width=4cm, height=4cm}}
 \endminipage\hfill
 \minipage{0.33\textwidth}%
 \centerline{\psfig{file=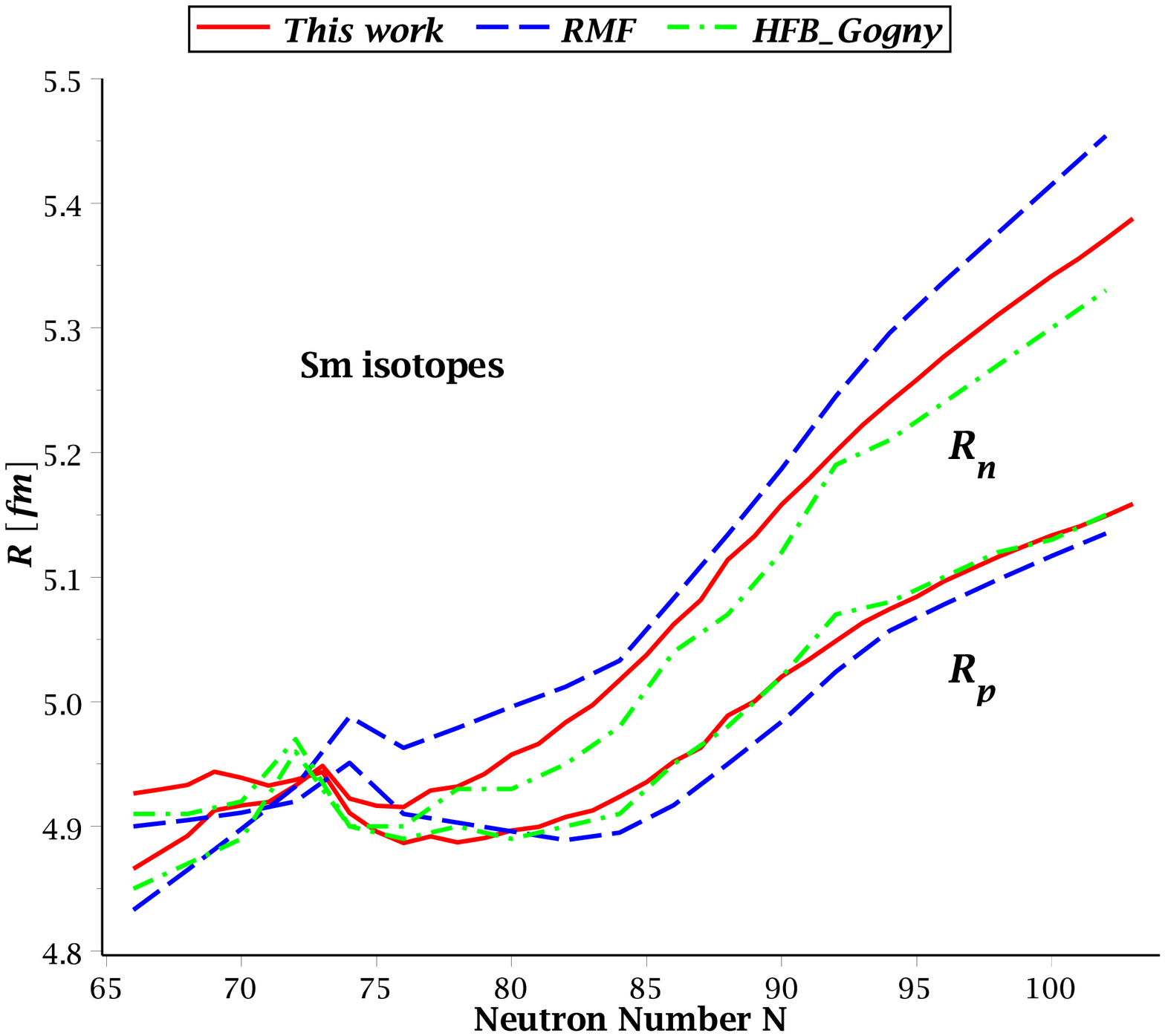,width=4cm, height=4cm}}
 \endminipage
 \caption{The neutron and proton radii of $Nd$, $Ce$ and $Sm$ isotopes.}
 \label{R}
 \end{figure}
 
Fig.\ref{R} shows the neutron and proton radii of $Nd$, $Ce$ and $Sm$ isotopes obtained in our calculations. The predictions of RMF theory and the HFB calculations based on the D1S Gogny force\cite{AMEDEE} are also given for comparison. we have plotted neutron and proton radii ($R_n$ and $R_p$) together in order to see the difference between them.

In the vicinity of the $\beta$-stability line ($N \approx  Z$), the neutron and proton radii are nearly  the same. But as the neutron number increases, the difference between the neutron and proton rms radii starts to increase in favour of developing a neutron skin.
This difference reaches $0.242$ fm for $^{161}$Nd, $0.257$ fm for $^{157}$Ce and $0.228$ fm for $^{165}$Sm, which can be considered as an indication of possible neutron halo in $Nd$, $Ce$ and $Sm$ isotopes. On the other hand, the neutron radii of the three isotopic chains show a kink about the neutron shell closure ($N=82$).\

\subsection{Quadrupole deformation}  
\begin{figure}[!htb]
 \minipage{0.33\textwidth}
 \centerline{\psfig{file=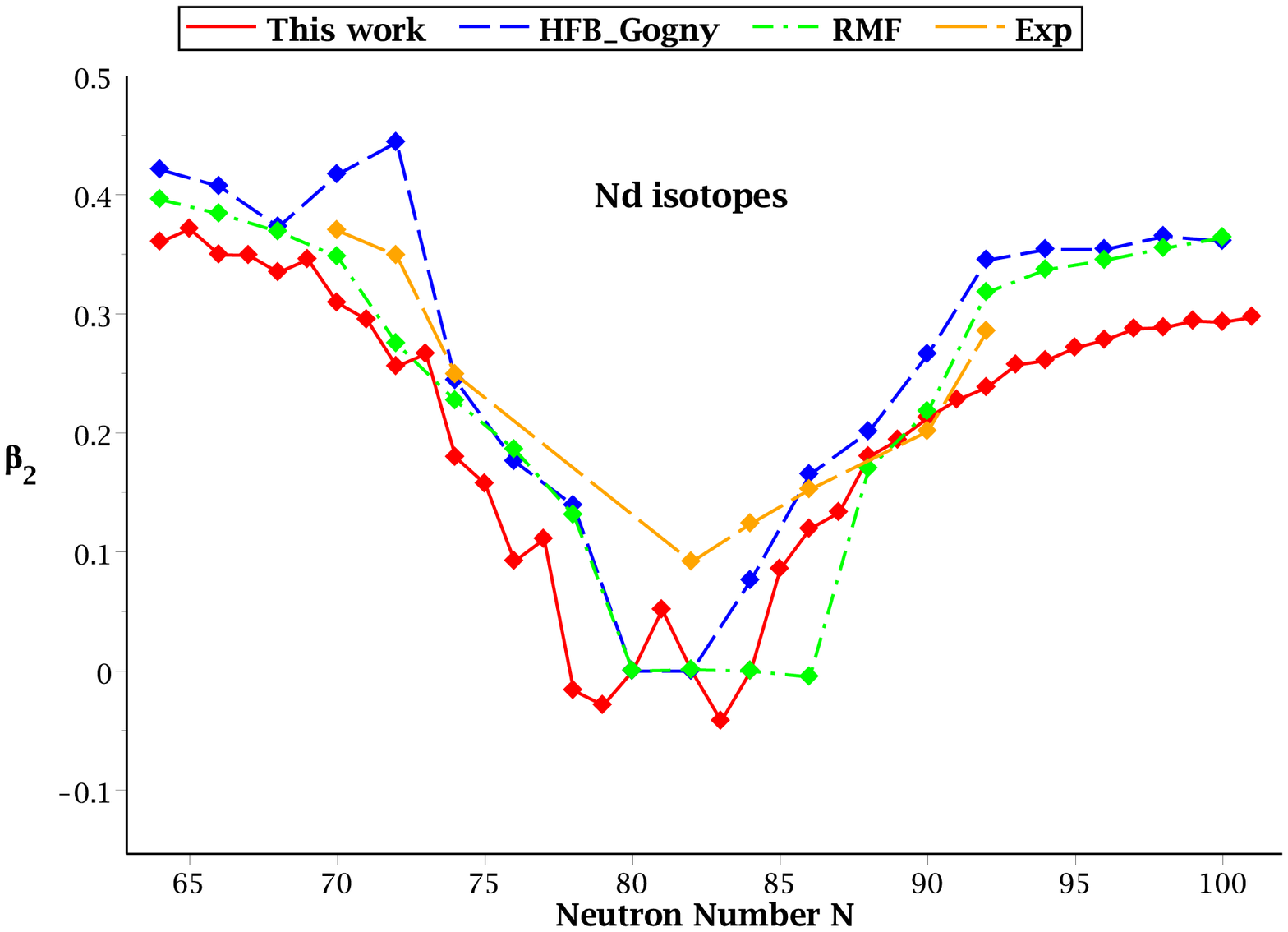,width=4cm, height=4cm}}
 \endminipage\hfill
 \minipage{0.33\textwidth}
 \centerline{\psfig{file=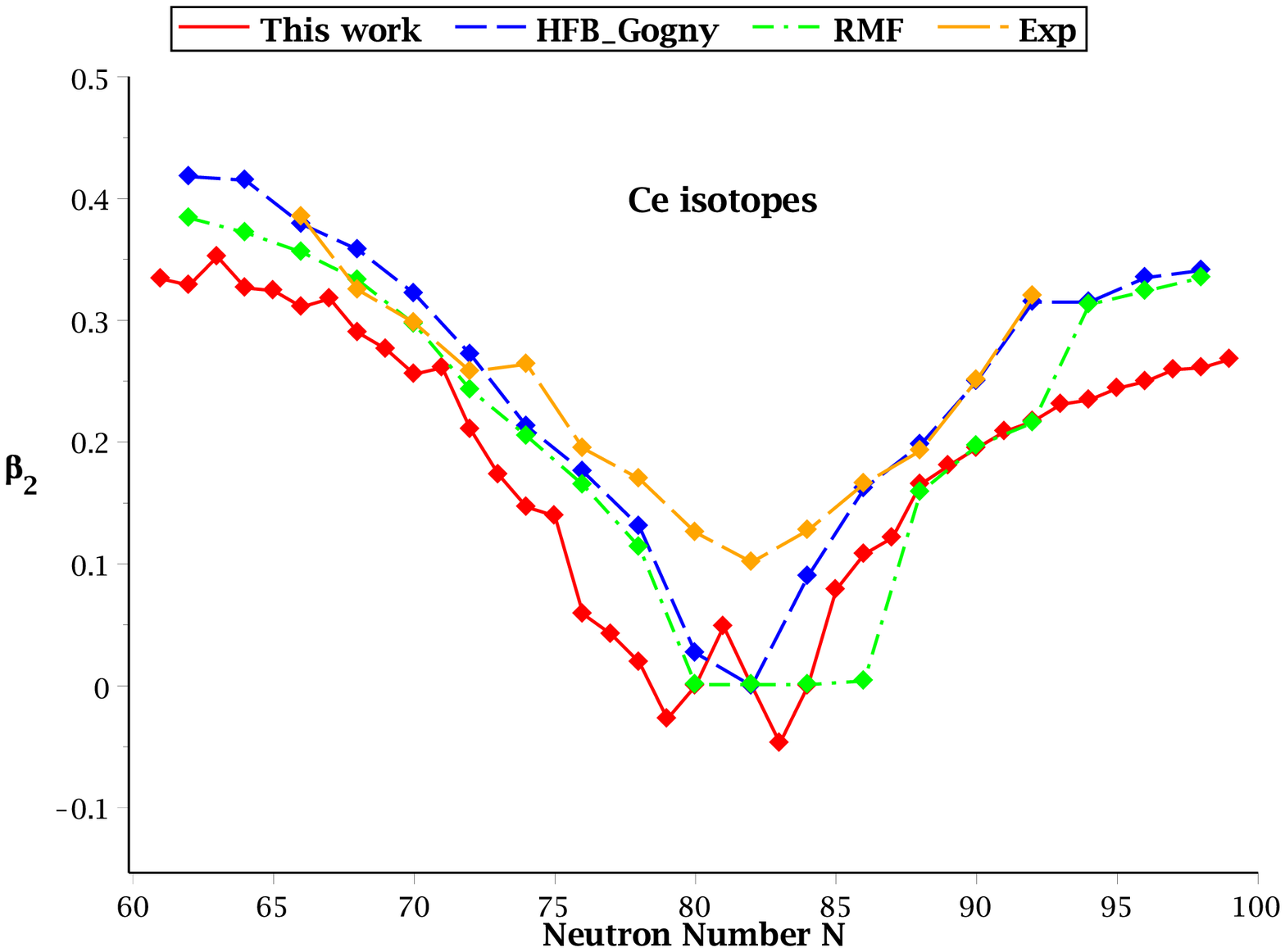,width=4cm, height=4cm}}
 \endminipage\hfill
 \minipage{0.33\textwidth}%
 \centerline{\psfig{file=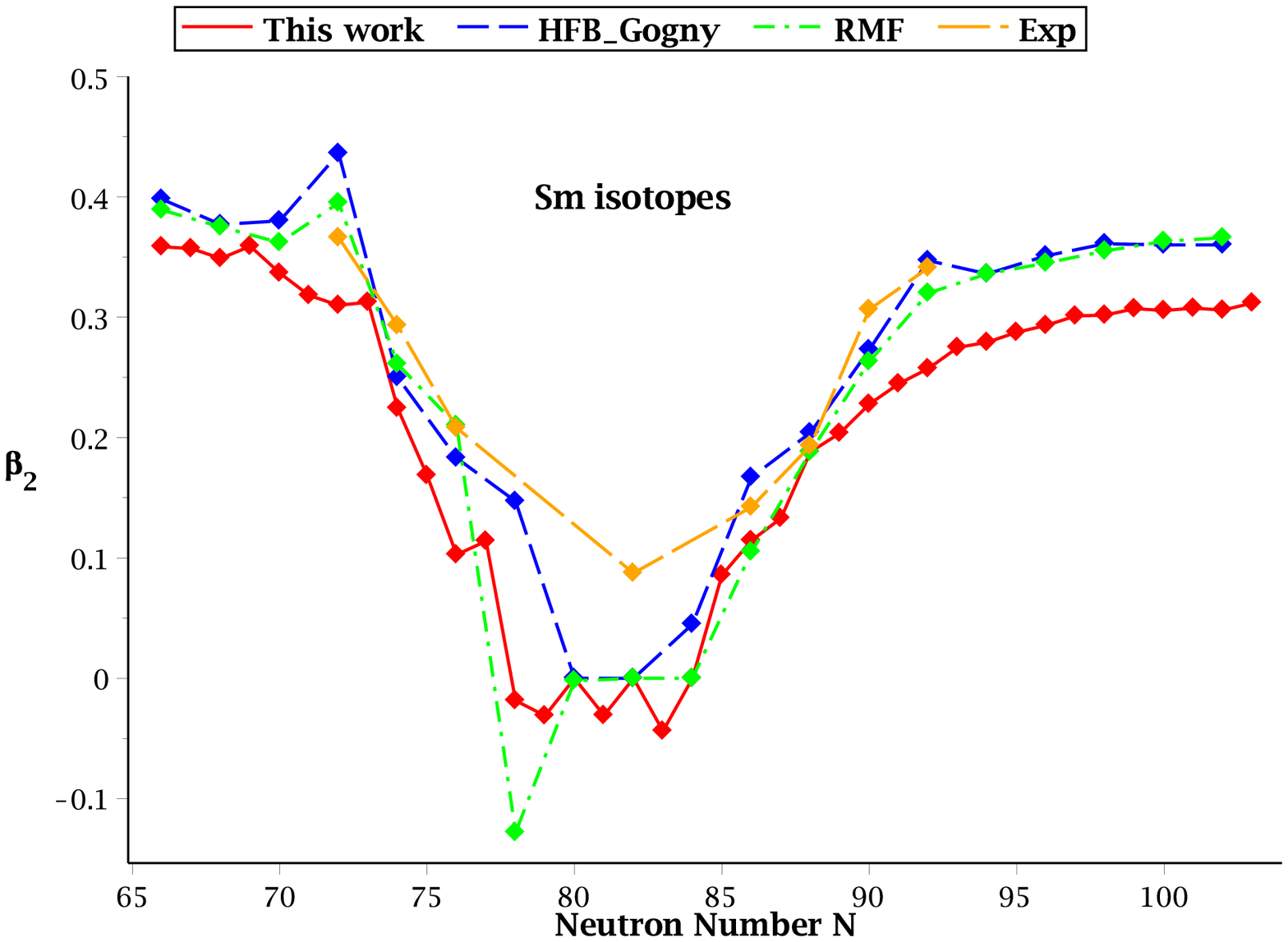,width=4cm, height=4cm}}
 \endminipage
 \caption{The Quadrupole deformation Parameters $\beta_2$ for Nd, Ce and Sm isotopes.}
 \label{beta}
 \end{figure}

The deformations of nuclei play a crucial role in determining their properties such as quadrupole moment, nuclear sizes and isotope shifts. In this subsection we compare the quadrupole deformation parameters $\beta_2$ obtained by our calculations with available experimental data\cite{beta_exp}. The $\beta_2$  values for the three isotopic chains considered in this work are shown in Fig.\ref{beta}. The results of the RMF \cite{Bayram2013} theory and the HFB calculations based on the D1S Gogny force\cite{AMEDEE} are also presented for comparison.

As it can be seen from Fig.\ref{beta}, the agreement between different theoretical calculations, this work, HFB based on the D1S Gogny force\cite{AMEDEE} and RMF \cite{Bayram2013} theory, and experimental data is quite good in general. The $\beta_2$ values show a minima at the magic neutron number $N=82$ as expected, because almost all nuclei with $N=82$ are spherical. Another well-known characteristic shown in Fig.\ref{beta} is that the even-odd nucleus tends to be deformed despite that its neighboring even-even nuclei are spherical. The $\beta_2$ values obtained in this work as well as those of RMF \cite{Bayram2013} theory, HFB calculations based on the D1S Gogny force\cite{AMEDEE} and available experimental data\cite{beta_exp} manifest an interesting change of shapes of nuclei below and above the magic neutron number $N = 82$. For nuclei below $N=82$, the three isotopic chains exhibit a transition from prolate shapes to spherical shapes, and for neutron number higher than $N=82$, the prolate deformation 
increases and then saturates at a value which closes to $\beta_2=0.29$, $\beta_2=0.30$ and $\beta_2=0.26$ for Nd, Sm and Ce isotopes, respectively.

Here it should be noted that shape coexistence and shape mixing phenomena are not investigated in the present work, because we limited our calculations to the axial frame. To study these features, more sophesticated beyond-mean-field models are used, such as the Generator Cordinate Method (GCM) \cite{Bender2014}, and the Bohr Hamiltonian \cite{oulne}.\\

\section{Conclusion}
HFB theory with Skyrme force SLy5 has been employed to investigate the ground-state properties of even-even and odd $Nd$, $Ce$ and $Sm$ isotopic chains. Binding energies, two-neutron separation energies, quadrupole deformation and proton, neutron and charge radii for $Nd$, $Ce$ and $Sm$ isotopes have been calculated. These calculations have been performed by means of a new generalized formula for pairing strength $V_0^{n,p}$ for neutrons and protons. The use of this formula has improved the results of ground-state properties of the nuclei.
The BE of $Nd$, $Ce$ and $Sm$ isotopes  have been described successfully in this work. The parabolic behavior of the BE/A has been  well reproduced in respect to the experimental curve.
A possible neutron halo has been observed in the three series of isotopes $Nd$, $Ce$ and $Sm$.
The kink related to isotopic shifts about the neutron-shell-closure $(N=82)$ was visible in our calculations, and the quadrupole deformation of Nd, Ce and Sm are well described with small differences from experimental results.


\begin{thebibliography}{0}
\bibitem{Navratil}P. Navratil, B.R. Barrett and W.E. Ormand, Phys. Rev C56, 2542 (1997).
\bibitem{Koonin}S. E. Koonin, D. J. Dean, and K. Langanke, Phys. Rep. 1, 278 (1997).
\bibitem{Dobaczewski}J. Dobaczewski, W. Nazarewicz, T.R. Werner, J.F.Berger, C.R. Chinn and J. Dechargé, Phys. Rev. C 53,
2809 (1996).
\bibitem{Terasaki} J. Terasaki, P.-H. Heenen, H. Flocard and P. Bonche, Nucl. Phys. A600, 371 (1996).

\bibitem{Stoitsov2000}M.V. Stoitsov, J. Dobaczewski, P. Ring and S. Pittel,Phys. Rev. C 61, 034311 (2000).

\bibitem{Chabanat} E. Chabanat, P. Bonche, P. Haensel, J. Meyer and R. Schaeffer, Nucl. Phys. A. 635, 231 (1998).

\bibitem{Teran} E. Ter\'{a}n, V.E. Oberacker, and A.S. Umar, Phys. Rev. C, 67, 064314 (2003).

\bibitem{Ring96} P. Ring, Prog. Part. Nucl. Phys. 37, 193 (1996).


\bibitem{Lalazissis}G. A. Lalazissis, S. Raman and P. Ring,{\it Atom. Data Nucl. Data Tables} 71, 1 (1999).

\bibitem{Werner}T.R. Werner, J.A. Sheikh, W. Nazarewicz, M.R. Strayer,
A.S. Umar, and M. Misu, Phys. Lett. B 333, 303 (1994).

\bibitem{Yamagami}M. Yamagami, K. Matsuyanagi, and M. Matsuo, Nucl.Phys. A693, 579 (2001).

\bibitem{Bender}M. Bender, P. H. Heenen and P. G. Reinhard, Rev. Mod. Phys. 75, 121 (2003).

\bibitem{Stoitsov}M. V. Stoitsov, J. Dobaczewski, W. Nazarewicz and P. Ring, Comp. Phys. Commun.
167, 43 (2005).

\bibitem{Ring} P. Ring and P. Schuck,{\it The Nuclear Many-Body Problem}, eds. W. Beiglböck et al.
(New York, Springer-Verlag, 1980). 

\bibitem{Greiner}W. Greiner, J. A. Maruhn,{\it Nuclear Models} Berlin, Springer-Verlag, (1995).

\bibitem{Stoitsov2013} M. V. Stoitsov, N. Schunck, M. Kortelainen, N. Michel, H. Nam, E. Olsen, S. Wild, Comp. Phys. Commun. 184, 1592 (2013).

\bibitem{Moller95} P. Moller, J. R. Nix, W. D. Myers, and W. J. Swiatecki. {\it Nuclear ground-state masses and deformations}. Atomic Data and Nuclear Data Tables, 59(2), 185-381,  (1995).

\bibitem{Dobaczewski2009} J. Dobaczewski, W. Satuła, B. G. Carlsson, J. Engel and al, Computer Physics Communications, 180(11), 2361-2391, (2009).

\bibitem{Bartel} J. Bartel, P. Quentin, M. Brack, C. Guet and H. B. Hakkansson, Nucl. Phys. A. 386, 183 (1982).

\bibitem{Baran} A. Baran, J. L. Egido, B. Nerlo-Pomorska, K. Pomorski, P. Ring and L. M. Robledo, J. Phys. G.
21, 657 (1995).

\bibitem{WANG} M. Wang, G. Audi, A. H. Wapstra and al. -{\it The Ame2012 atomic mass evaluation}. Chinese Physics C, 36, 1603 (2012).

\bibitem{AMEDEE} \textit{$http://www-phynu.cea.fr/HFB-Gogny_eng.htm$}.

\bibitem{Article} S. O. Kara, T. Bayram, S. Akkoyun, Journal of Physics: Conference Series 490 012106, (2014).

\bibitem{Moller97}P. M\"{o}ller, J. R. Nix and K.-L. Kratz, Atom.{\it Data Nucl. Data Tables} 66, 131 (1997).

\bibitem{Angeli}I. Angeli. {\it Atomic data and nuclear data tables}, 87(2), 185-206 (2004).



\bibitem{beta_exp}https://www-nds.iaea.org/RIPL-2/masses/gs-deformations-exp.dat (30/07/2015).

\bibitem{Bayram2013} T. Bayram and A. H. Yilmaz,{\it Table of Ground State Properties of Nuclei in the RMF Model}. Modern Physics Letters A, 28(16), 1350068  (2013).

\bibitem{Bender2014} B. Bally, B. Avez, M. Bender, and P. H. Heenen, Physical review letters, 113(16), 162501, (2014).

\bibitem{oulne} M. Chabab, A. Lahbas and M. Oulne, Phys. Rev. C, 91, 064307 (2015).

\end{thebibliography}
\end{document}